**Methyl cellulose/cellulose nanocrystal nanocomposite fibers with high ductility**

Ville Hynninen[a], Pezhman Mohammadi[b], Wolfgang Wagermaier[c], Sami Hietala[d], Markus B. Linder[b], Olli Ikkala[a,b*], Nonappa[a,b*]

[a] Department of Applied Physics, School of Science, Aalto University, P.O. Box 15100, FI-00076 Espoo, Finland

[b] Department of Bioproducts and Biosystems, School of Chemical Engineering, Aalto University, P.O. Box 16300, FI-00076 Espoo, Finland

[c] Department of Biomaterials, Max Planck Institute of Colloids and Interfaces, D-14424 Potsdam, Germany

[d] Department of Chemistry, University of Helsinki, P.O. Box 55, FI-00014 HY Helsinki, Finland

**ABSTRACT**

Methylcellulose/cellulose nanocrystal (MC/CNC) nanocomposite fibers showing high ductility and high modulus of toughness were prepared by a simple aqueous wet-spinning from corresponding nanocomposite hydrogels into ethanol coagulation bath followed by drying. The hydrogel MC aq. concentration was maintained at 1 wt-% while the CNC aq. loading was systematically varied in the range 0 – 3 wt-%. This approach resulted in MC/CNC fiber compositions from 25/75 wt-%/wt-% to 95/5 wt-%/wt-%. The optimal mechanical properties were achieved with the MC/CNC composition of 80/20 wt-%/wt-% allowing high strain (36.1 %) and modulus of toughness (48.3 MJ/m$^3$), still keeping a high strength (190 MPa). Further, we demonstrate that the continuous spinning of MC/CNC fibers is potentially possible. The results indicate possibilities to spin MC-based highly ductile composite fibers from environmentally benign aqueous solvents.

# 1. Introduction

Because of the natural abundance, biocompatibility, and biodegradability, cellulose and its derivatives have extensively been studied and utilized for a wide variety of applications, including tissue engineering, regenerative medicine, nanocomposites, fibers, membranes, and photonics.[1] The possibilities to tune the cellulose chemistry allow to control the underlying physicochemical properties efficiently.[2] Among the cellulose derivatives, methyl cellulose (MC) has extensively been used for applications related to foods, detergents, paints, adhesives, cosmetics, safety, and gels.[3–5] Therein, MC acts as an emulsifier, viscosity controller, and mechanical adjuvant for composite systems.[6] It has been shown that 2.49 wt% of MC having molecular weight of 310 kDa and a degree of substitution of 1.8 display lower critical solution temperature (LCST) at 29 ± 2 °C, below which Newtonian flow is observed and above which non-Newtonian flow and increased viscosity emerge.[7] The aqueous dispersions exhibit thermosensitive hydrogelation.[8–17] The gelation mechanisms are subtle, depending on the molecular weight, degree of substitution (DS), temperature, and shear rate.[13,16–24] The stiffness of an MC hydrogel can reversibly be increased through relatively mild heating to ~60 °C.[14,17,25] The above-mentioned properties along with its biodegradable nature, MC has been studied as a potential alternative to oil-based polymers, such as composite gels with poly(caprolactone), cement pastes, and composites with hydroxyapatite.[26–28]

On the other hand, interest on nanocellulose composites has been growing rapidly in recent years.[29] Among nanocelluloses, cellulose nanocrystals (CNCs) are attractive reinforcing agents because of their large specific surface area, excellent mechanical properties (elastic modulus ~150 GPa), lightweight (density ~1.566 g/cm$^3$), sustainability, and renewability.[29,30] Depending on their surface functionalization and the nature of the polymer matrix, the mechanical properties of the nanocomposites can be tuned based on the interactions and percolation within the matrix. Composite films with CNC have been shown using polylactic acid (PLA), polycaprolactone, and poly(vinyl alcohol) (PVA).[31–33] MC nanocomposite films with CNC show improved thermomechanical and barriers properties.[34] CNCs modulate the gelation properties of polysaccharides.[35] The MC/CNC hybrids have recently been used as stabilizers for nanocomposite latexes.[36,37] The MC/CNC ratios allow tuning of the thermoreversible hydrogelation and the moduli.[38]. Another characteristic property of CNCs is that their aqueous dispersions show lyotropic chiral nematic liquid crystallinity above the critical concentration C* ≈ 4 wt%.[39–42] Based on that, the MC/CNC composite hydrogels also show birefringence due to liquid crystallinity, interestingly, emerging at a low aq. CNC loading of 1.0 – 1.5 wt-% with the fixed MC aq. concentration of 1.0 wt-%.[43] Within the MC/CNC gels, they allow enhanced interactions by entanglements of the MC polymers and CNCs, thus providing physical cross-linking sites and making the gel network stronger (Fig. 1d-f). This property is effectively enhanced by the low overlap concentration of MC that is in the range of 0.03 × 10$^{-2}$ g/mL.[44] Therefore, MC links the CNCs together, and regulates the organization and malleability of the gel.

The hybrids of MC and CNC suggest fundamentally interesting materials due to their biocompatibility and suggested safety.[9,45] The possibilities to prepare MC/CNC composite aqueous gels at ambient conditions at room temperature at very low CNC loadings along with its shear thinning properties suggest to consider whether such properties can be exploited to aqueous fiber spinning by extrusion. In more general, explorations of fibers consisting of stiff and strong elements embedded in a soft and energy dissipating matrix are also encouraged by spider silk.[46,47] In the present approach, the rigid CNCs would be captured within the amorphous MC network. Therein, the CNCs potentially behave as quasi-mobile anchorage points reinforcing the material through favorable interactions with MC and their extended dissipative potential due to the ability to re-organize and move in the softer MC matrix.[48,49]

In retrospect, cellulosic fibers are well established and the current commercially available regenerated cellulose fibers exhibit very good and broadly tunable mechanical properties, involving considerable economic impact.[50] In addition to the textile industry, they are precursors for carbon fibers and as reinforcing components in composite materials.[51] However, regenerated cellulose fibers are generally prepared through a dissolution of a cellulose source material in rather harsh solvents followed by spinning. For example, in a typical viscose process cellulose is treated with sodium hydroxide and carbon disulfide before being spun into an acidic precipitation solvent. The more recent and greener Lyocell process relies on a direct dissolution of cellulose into an aqueous solution of *N*-methyl morpholine-*N*-oxide prior to spinning. Recently, ionic liquids have also been studied as a more potential and alternative route to decrease possible chemical and process related risks.[52] Thus it is encouraged to search for more environmentally friendly aqueous processes for cellulose fiber spinning.

In the following, cellulosic nanocomposite fibers of MC and CNCs from aqueous medium are described. Importantly, pure MC or CNC alone did not allow solid fibers, whereas the combination of MC and CNC appeared critical for fiber spinning. We show fibers with high ductility accompanied with relatively good strength.

**2. Experimental**

**2.1 Materials**

Methylcellulose (MC, MW 88,000, product no. M0512, Lot# 079K0054V) was acquired from Sigma-Aldrich. MC had a methoxy substitution of 27.5 – 31.5 % (weight) and a degree of substitution of 1.5 – 1.9 as reported by the supplier. For CNC preparation Whatman® Grade 1 qualitative filter papers (cat no. 1001 125) and Whatman® Grade 541 hardened ashless filter papers (cat no. 1541-125, Lot# 9722517) and Spectra/Por® 1 standard regenerated cellulose dialysis tubing with molecular weight cut-off of 6 – 8 kDa (part no. 132665, Lot# 9200679, Spectrum Laboratories, Inc.) were purchased from VWR. 96.1 % (v/v) Ethanol (Etax A, Altia Inc.) was used for fiber spinning coagulation bath. Other chemicals were obtained from Sigma-Aldrich and used as received. Ultrapure MilliQ® H$_2$O (18 mΩ) was used in all experiments.

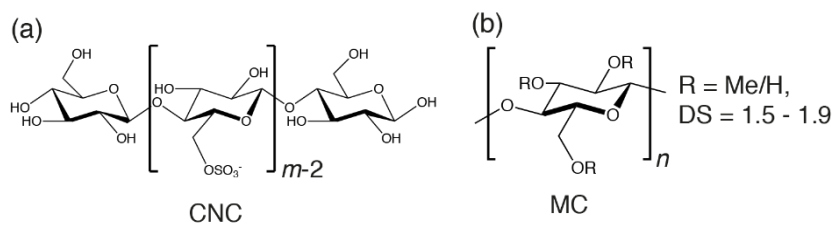

**Scheme 1:** Chemical structures of cellulose nanocrystal nanorods and methylcellulose polymers.

## 2.2 Preparation of CNCs

CNCs were prepared according to a previously published procedure.[53] In short, Whatman® Grade 1 filter papers were mechanically ground into homogenous powder. The powder was hydrolyzed with 64 % sulfuric acid under gentle stirring (32 rpm) at 45 °C for 45 min. Typically, 272.3 g of sulfuric acid (95 – 97 %) was mixed with 136.2 g of MilliQ® H$_2$O, and the mixture was used to treat 15 g of filter paper powder. The reaction was quenched by diluting approximately 10-fold by adding 3 L of MQ® H$_2$O and then allowed to sediment undisturbed for at least 20 hours. The clear supernatant was discarded and the remaining cellulose dispersion washed by two rounds of centrifugation (Wifug X-3 centrifuge with a fixed angle rotor): First, at 6000 rpm for 20 min and then 2500 rpm for 45 min, both at ambient temperature. After each centrifugation, the clear supernatant was discarded and the pellet redispersed in MQ® H$_2$O by mechanical agitation. The resulting CNCs were further purified by dialysis against MQ® H$_2$O until the conductivity of the dialysate remained below 5 $\mu$S/cm. Finally, the CNC dispersion was filtered through Whatman® 541 filter paper and stored at + 4 °C until use. 3.47 % (34.7 mg/mL) CNC stock solution was prepared through controlled evaporation of water at 45 °C and under constant stirring (250 rpm). The solid content of the CNC concentration was determined gravimetrically and the material was further characterized by TEM, dynamic light scattering, zeta potential measurement, and conductometric titration.

## 2.3 Preparation of MC-CNC hydrogels

Hydrogels of MC-CNC with a fixed aqueous MC content of 1 wt-% and a varying CNC aqueous loading of 0 – 3.0 wt-%, were prepared. As the solubilization of MC is subtle, the supplier's instructions to properly disperse MC were followed: First, the appropriate amount of CNC stock solution to reach the desired final concentration and 1/3 of the required volume of MQ® $H_2O$, that had been heated up to 85 °C, were added to the solid MC. The dispersion was agitated until all MC had become wetted. The rest of the water was then added under continuous stirring as cold water and the mixture kept under stirring for at least 15 min to ensure thorough mixing of components. Thereafter, the mixture was cooled to 4 °C for 1 hour and finally further stirred for at least 40 h at ambient temperature (22 °C). The ready-made mixtures were stored at 4 °C in a refrigerator until use.

## 2.4 Fiber spinning

MC/CNC hydrogel was spun into a continuous fiber by using a previously described extrusion system consisting of a high-pressure pump, a sample container (V = 10 mL), an extrusion capillary tube (L = 1.50 m, Ø = 0.5 mm), and a coagulation bath containing 96.1 % (v/v) ethanol (Fig. S1).[54] Volumetric extrusion flow rate of 1.0 mL/min, i.e. linear flow of 509.3 cm/min, was used (Video S1). Before spinning, the MC-CNC material was centrifuged at 3000 rpm for 1 – 5 min to remove air bubbles. After the extrusion, the fibers were allowed to equilibrate in the ethanol bath for 10 min before they were cut to approximately 8 cm long pieces and hung to dry at ambient conditions fixed at both ends for at least 20 h. The procedure was repeated for all sample compositions. Additionally, scaled-up continuous spinning process was shown possible by using the same equipment complemented with a rotating collector arm, by tuning the extrusion flow rate to 0.06 mL/min (33.95 cm/min) to match the rotating collector winding rate and to maximize the dwell time within the coagulation bath regardless of the constant motion, and by replacing ethanol with isopropanol to facilitate coagulation (see Video S2). Table S1 shows the summary of all used compositions of the MC/CNC gels, corresponding relative compositions of the dried fibers, and the average mechanical properties of the fibers.

## 2.5 Mechanical characterization

Uniaxial tensile tests were performed with a 5 kN tensile/compression module (Kammrath & Weiss GmpH, Germany) equipped with a 100 N load cell and operating inside a controlled humidity box. The elongation speed was 8.35 $\mu$m/s and the gauge length 1.0 cm (Video S3). Before the testing, the ends of the fibers were fixed by gluing them between two adhesive paper pieces cut from Staples Stickies™ sticky notes and the samples were equilibrated at 50 % relative humidity inside the humidity box for at least 20 h. Additionally, the test was repeated at 20 % and 78 % relative humidities for the mechanically best fiber composition to monitor the effect of different humidity conditions. The data was analyzed with Matlab (MathWorks) to obtain the ultimate strength, ultimate strain, Young's modulus and modulus of toughness as described earlier.[54] Ultimate strength and ultimate strain values were determined as the highest stress and the maximum elongation reached by each sample before breaking, respectively. The Young's modulus was defined as the average slope of the of the stress-strain curve's elastic region prior to the yield point, and the modulus of toughness as the area remaining under the stress-strain curve. The fiber cross-sectional areas required in the analysis were determined from SEM images of unstretched fibers by using ImageJ.[55] The SEM samples pieces were cut from each tensile test specimen during the sample preparation by a razor blade.

### 2.6 Rheology

Rheological characterization of MC-CNC hydrogels was performed with TA AR2000 stress-controlled rheometer equipped with 20 mm steel plate-plate geometry at 20 °C. Flow properties of mixtures with MC/CNC aqueous concentrations 1.0/0.0, 1.0/0.25, 1.0/1.5, and 1.0/3.0 wt-%/wt-% were studied in the shear rate range of $0.1 - 1000$ s$^{-1}$.

### 2.7 Polarized optical microscopy (POM)

The birefringence of the fibers was qualitatively assessed by POM. Samples placed between two microscope glasses were inspected through crossed polarizers with Leica DM4500 P polarization microscope combined with Canon EOS 80D DSLR camera at 22 °C.

### 2.8 Scanning electron microscopy (SEM)

SEM imaging was done with Zeiss Sigma VP scanning electron microscope at 1.5 kV acceleration voltage. Fiber samples were attached on aluminum SEM stubs with double-sided carbon tape and sputtered with 6 nm thick platinum-palladium coating prior to imaging by using Leica EM ACE600 high vacuum sputter coater. For the cross-sectional area imaging, an additional 5 mm thick aluminum block was affixed onto the standard SEM stub and the fibers were fastened upright to the sides of the block to allow direct cross-sectional view. MC-CNC gel samples were imaged as freeze-dried aerogels. Approximately 300 $\mu$l of gel was placed inside a 2.0 mL centrifuge tube, frozen by plunging into liquid nitrogen for 5 min and then freeze-dried in a lyophilizer (0.840 mbar, - 100 °C) for 44 h. The resulting aerogel was attached on a standard aluminum SEM stub with double-sided carbon tape and sputtered with 6 nm of platinum-palladium before imaging.

### 2.9 Wide angle X-ray scattering (WAXS)

Wide-angle X-ray diffraction experiments were carried out at the µSpot beamline at BESSY II synchrotron source (Berliner Elektronenspeicherring-Gesellschaft für Synchrotronstrahlung, Helmholtz-Zentrum Berlin, Germany). Collection of the diffraction patterns was performed using a multilayer silicon (111) monochromator with an X-ray wavelength of 0.82656Å (energy of 15 keV) and a beam size of 50 μm. Calibration was performed with a quartz (SiO$_4$) with sample to detector distance of around 280mm. Data was collected on a two-dimensional CCD detector (EIGER X 9M, with an area of 233 by 245 mm$^2$) with a pixel size of 75 μm$^2$. Two fibers for each specimen were clamped in a sample holder in order to position the samples perpendicular to the beam path. Three positions along the length of the fibers were measured. After subtraction of air scattering from the diffractogram, azimuthal intensity profiles at the (004) reflection extracted by sector-wise integration after masking the diffractogram to show only the corresponding diffraction peak. The base line for azimuthal intensity profile adjusted to zero to calculate Hermans orientation parameter (S) according to equations 1, where Φ is the azimuthal angle.[54,56]

$$S = \frac{3}{2}\langle cos^2\Phi\rangle - \frac{1}{2}$$

(1)

Where the mean-square cosine is calculated from I(Φ) the scattered intensity by integrating over the azimuthal angle Φ as shown in equation (2).

$$\langle cos^2\Phi\rangle = \frac{\sum_0^\pi I(\Phi)sin\Phi cos^2\Phi}{\sum^\pi I(\Phi)sin\Phi}$$

(2)

Depending on the orientation, the values of S can be 1 (perfect vertical orientation), 0 (for isotropic orientation) and -0.5 (horizontal orientation).

**2.10 Conductometric titration**

Conductometric titration was performed with 751 GPD Titrino (Methrom AG) conductometric titrator together with Tiamo software as described earlier and in SCAN-CM 65:02 standard procedure.[57,58] Prior to the titration, the acidic groups on CNC surfaces were protonated by adding concentrated hydrochloric acid (HCl) to the CNC dispersion to the final concentration of 0.1 M. After 15 min of equilibration at 22 °C, the excess acid was removed by dialysis against MQ® H$_2$O until the conductivity of the dialysate remained below 5 $\mu$S/cm. For the titration, 20 mL of the protonated CNC dispersion (c = 15.9 mg/mL) was mixed with 490 mL of degassed MQ® H$_2$O, 0.5 mL of 0.1 M HCl and 1.0 mL of 0.5M NaCl. The resulting dispersion was titrated against 0.1 M NaOH under constant stirring of 300 rpm. The titrant was added in 0.02 mL increments every 30 s. The acidic sulfate half ester content was calculated and reported as the ratio of the amount of titrant needed to neutralize the acidic sulfate groups in $\mu$mol to the amount of CNC (g) in the dispersion.

**2.11 Transmission electron microscopy (TEM)**

TEM was used to determine the CNC size distribution. CNC stock solution was first diluted to 0.1 mg/mL and 0.05 mg/mL dispersions by adding MQ® $H_2O$. 10 µL droplet of a diluted CNC dispersion was pipetted onto a plasma-cleaned (30 s, Gatan Solarus 950) copper TEM grid with ultra thin carbon only support film (CF200-Cu-UL, Lot# 171012, Electron Microscopy Sciences) and allowed to absorb for 1 min. The excess material was blotted with filter paper. The samples were imaged with JEM-2800 (JEOL) high resolution TEM operating at 200 kV and analyzed with ImageJ software package.[55]

**2.12 Dynamic light scattering (DLS) and zeta potential (ζ) characterization of CNCs**

The hydrodynamic size distribution and the zeta (ζ) potential distribution were determined by using Zetasizer Nano ZS90 (Malvern Instruments). 12 mm square polystyrene cuvettes (Product no. DTS0012, Malvern Instruments) were used for DLS, and folded capillary zeta cell cuvettes (Product no. DTS1070, Malvern Instruments) for zeta potential measurements. Mixtures with the aq. MC/CNC ratios of 0.1/0.0, 0.0/0.1, and 0.1/0.1 wt-%/wt-% were analyzed. For the zeta potential measurement, NaCl was added to each sample to the final concentration of 1 mM. The samples were prepared as described above for MC-CNC mixtures. The reported values are the average of at least three measurements.

**3. Results and discussion**

**3. 1 Characterization of CNCs and MC/CNC hydrogel dopes**

The morphology, size distribution, and surface charge of CNCs were studied using TEM, DLS, and conductometric titration, respectively. The average length determined from the TEM image analysis was 238 ± 87 nm and the average aspect ratio 16, which is in agreement with the typical size of CNCs reported in the literature (Fig. 1a-b and Fig. S2).[59] Conductometric titration revealed the acidic sulfate half ester content to be 239.0 µmol/g (Fig. S3), which also agrees with the earlier reports.[60] The zeta potential (ξ) was – 48.8 mV suggesting excellent colloidal stability through Coulombic interactions (Fig. S4a). Pure CNC dispersion behaved like a free-flowing viscous liquid at the used concentration range (below aq. 3.4 wt-%) (Fig. 1c). At these concentrations, the aq. dispersion of pure CNC did not display birefringence under POM (Fig. S5), as expected.[42,61]

Upon mixing MC and CNC, the viscosity increased. Depending on the CNC concentration, gelation could be achieved at room temperature, as shown by the vial inversion tests (Fig. 1c). Consistent with earlier reports, morphologically the composite gel structure became more connected and sheet-like and less fibrillar upon increasing CNC concentration, as observed by SEM analysis (Fig. S6).[43] More quantitatively, the flow behavior of the aqueous mixtures was characterized by rheology where shear thinning was observed, a property that is beneficial for the fiber spinning process (Fig. S7). Recently, extensive rheological characterization of similar MC/CNC hydrogels have been performed by us that have revealed the storage modulus ($G'$) to be tunable in the range of approximately 1 – 100 Pa only by adjusting the aq. CNC loading between 0 wt-% and 3.5 wt-% while keeping MC concentration fixed at aq. 1 wt-%.[38,43] The connectivity between the CNC and MC, manifested by the gelation, turned relevant in the following to allow ductile nanocomposite fibers (Fig. 1d-f).

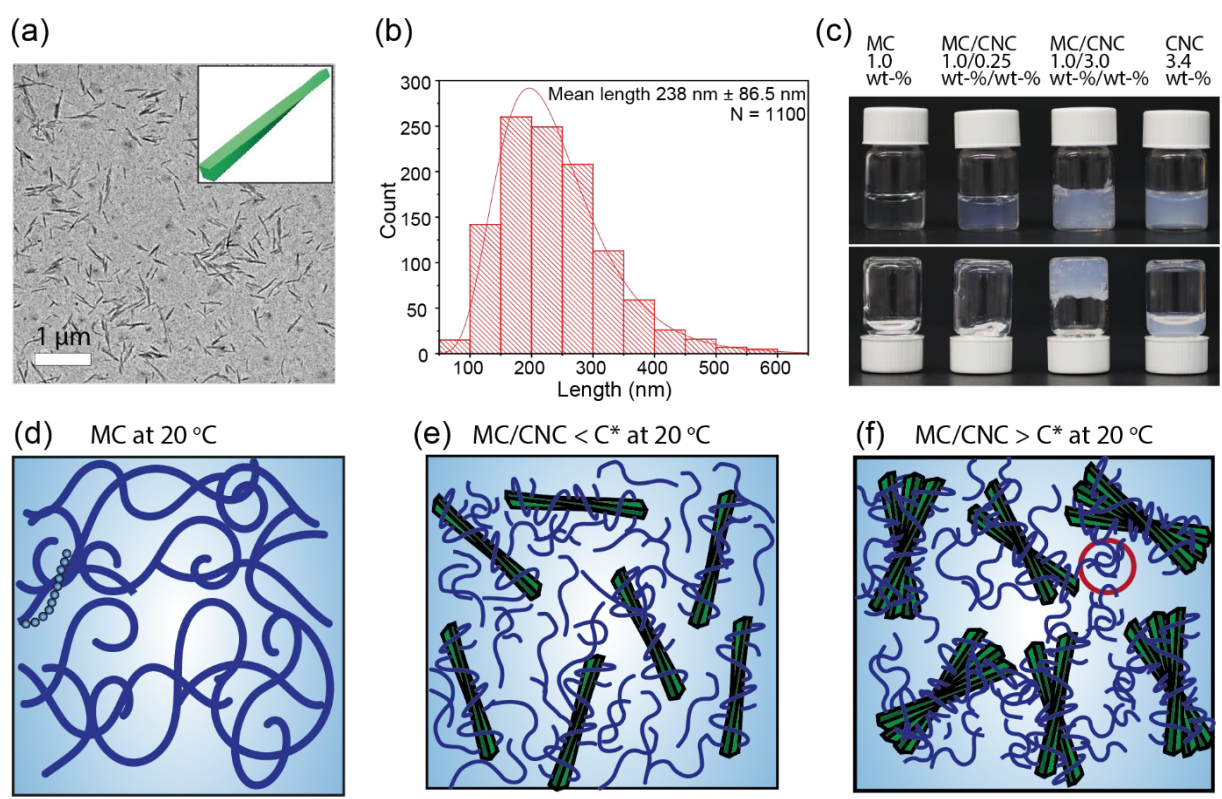

**Fig. 1.** MC and CNC materials. (a) TEM image of 0.1 mg/mL aq. dispersion of CNCs. Structural right-handed chirality of CNC rods is schematically shown in the inset. (b) Size distribution of aq. CNC particles determined from TEM images. (c) Vial inversion test indicating qualitatively the gelation. (d) Schematic representation of methyl cellulose polymer chains. Note that in the present compositions the MC concentration exceeds the overlap concentration, which is also emphasized with red circle in (f). (e and f) Schematic representation of MC/CNC composite hydrogels below and above the critical liquid crystalline concentration of the aq. hybrids (C*), manifesting birefringence.

MC and CNC have been shown to interact through enthalpic mechanisms in aqueous media at 25 °C.[43] Also, fast absorption of MC on CNC films has been demonstrated.[37] The favorable interaction was indirectly observed here by DLS, upon studying the dispersions at low concentrations (Fig. S4b). An accurate evaluation of the hydrodynamic size of CNCs by DLS is challenging due to the rod-like shape, but qualitatively the interaction between the two components and the resulting growth in particle sizes could be monitored. Pure aq. CNC gave a well-defined signal, whereas pure aq. MC resulted in a set of random peaks. By contrast, aq. MC/CNC 0.1/0.1 wt-%/wt-% gave a nearly identical signal to that of pure aq. CNC, but the peaks had shifted towards larger hydrodynamic size. Therefore, the MC is concluded to have been attached and conformed onto the CNCs, while retaining the colloidal distribution of CNCs. The zeta potential of the mixture was only slightly negative, - 2.0 mV, suggesting significantly lowered electrostatic stability compared to – 48.8 mV of pure CNCs (Fig. S4a). Thus, in the mixtures MC/CNC, the electrostatic stabilization of CNCs is suggested to be overtaken by steric stabilization originating from the MC chains. In comparison, the zeta potential of pure 0.1 wt-% MC was nearly neutral, - 1.7 mV.

### 3.2 Morphology of the pristine MC-CNC fibers

The composite MC/CNC fibers were prepared from the MC-CNC hydrogels upon extrusion into an ethanol coagulation bath using a fixed MC aq. concentration of 1 wt-% and a varying the aq. CNC loading from 0.0 to 3.0 wt-%, see Table S1 for all used compositions. Thus, the relative compositions in the dried fibers varied from MC/CNC 100/0.0 wt-%/wt-% (prepared from MC/CNC 1.0/0.0 wt-%/wt-% hydrogel) to MC/CNC 25/75 wt-%/wt-% (from MC/CNC 1.0/3.0 wt-%/wt-% hydrogel) based on the mass of the components. Note that neither pure MC nor pure CNC alone produced any fibers, but the combination of the two components was required. The CNC upper aq. loading was limited to 3.0 wt-% due to increased viscosity that caused excessive backpressure buildup in the extrusion pump and hindered the extrusion process at higher concentrations. After drying at ambient conditions, the fibers were characterized by POM, SEM and tensile tests to find out the optimal compositions for the best mechanical properties.

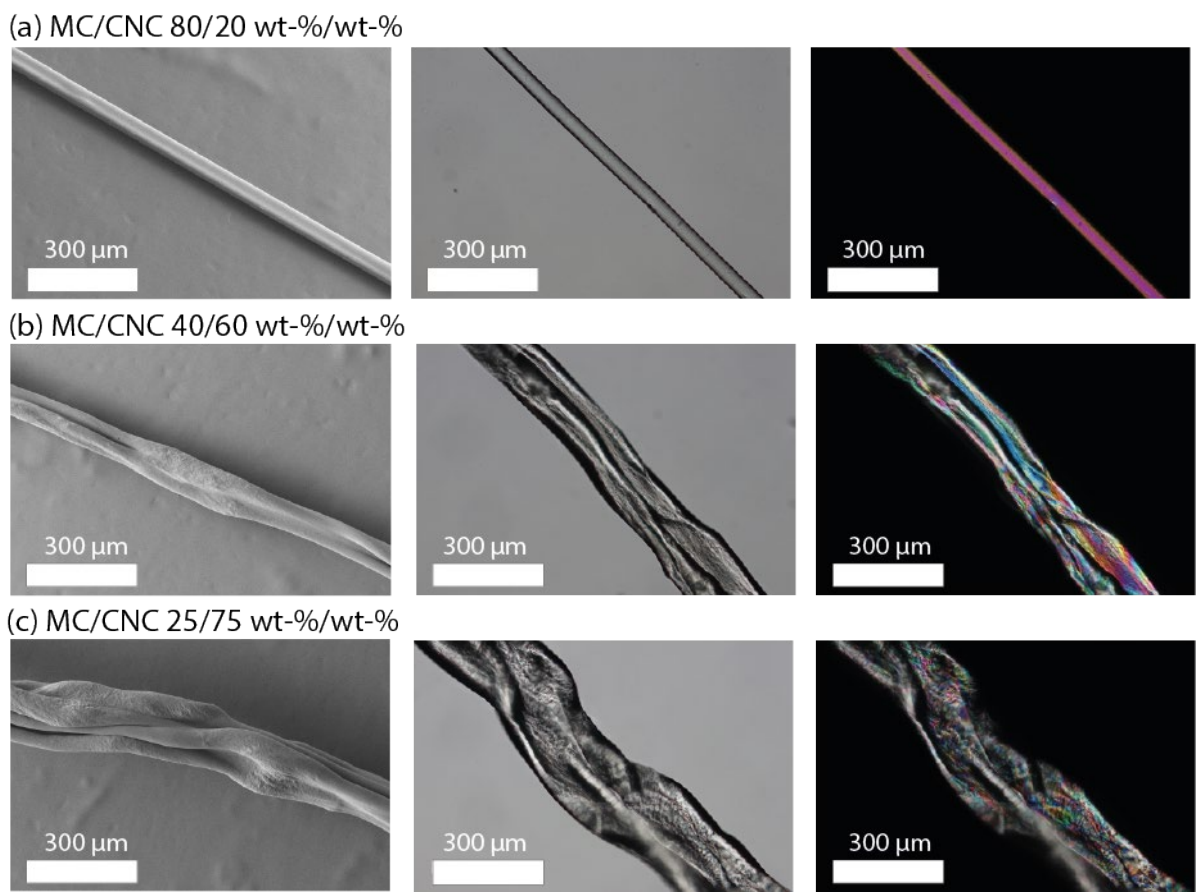

**Fig. 2.** Morphology of MC/CNC fibers based on SEM, light microscopy, and POM micrographs for different compositions. Increase in twisting and buckling is obvious with increasing CNC loading.

POM and SEM micrographs of the fibers with different compositions are shown in Figure 2. Typically, the fibers with low CNC loading appeared smooth and straight, whereas higher CNC loading correlated with rough and increasingly buckled surfaces. In all cases, the fibers were nearly transparent and appeared colorless to the naked eye. The CNC concentration was also reflected in the behavior of the fibers already in the wet state during the spinning process, where the MC/CNC mixtures with low CNC loading showed visible lateral and sheet-like spreading in the coagulation bath (Fig. S8). Such fibers expressed often flat and folded cross-sections due to the spinning and drying process while retaining smooth surfaces (Fig. 2a, Fig. S9a and Fig. S10). By contrast, fibers with high CNC loading did not show noticeable macroscopic spreading in the coagulation bath and retained their extruded form. Nonetheless, such fibers typically showed rough and extensively buckled surfaces and the fiber diameter increased with the increasing CNC loading (Fig. 2b-c, Fig. S9-S11). The low CNC fibers expressed homogeneous and continuous birefringence in POM (Fig. 2a) whereas the high CNC fibers showed more patched colorful patterns with decreasing long-range uniformity, qualitatively suggesting the existence of locally oriented liquid crystalline-like domains within high-CNC-loaded fibers. Therefore, the CNCs were hypothesized to assemble into liquid crystalline-like local regions (tactoids) inside the fibers (illustrated for the hydrogels in Fig. 1e-f), which then cascade into the rough and twisted features noticeable both at micro- and macroscale. The observed local liquid crystalline patterns in POM were reminiscent of the heterogeneously arranged phase-separated CNC liquid crystal mesophases within sodium sulfate-coagulated poly(vinyl alcohol)-CNC composite fibers reported previously.[62]

## 3.3 Mechanical characterization of the fibers

The tensile stress-strain curves of MC/NC fibers are shown in Fig. 3 at the relative humidity of 50 %. Not surprisingly, pure CNC did not allow fiber formation. The relative composition MC/CNC 25/75 wt-%/wt-% showed a brittle behavior without any observable plastic deformation (Fig. 3 top). Upon increasing the relative MC content, plastic deformations emerged, manifesting a yield strength at ca. 70 MPa and increased strains. For example, MC/CNC 50/50 wt-%/wt-% showed a clear yield and ultimate strain of 7 %. Upon further increasing the MC fraction to MC/CNC 80/20 wt-%/wt-%, higher ultimate strain was achieved together with increased ultimate strength. Finally, pure MC did not allow fiber spinning in the present setup. Fig. 4 shows the characteristic mechanical properties as a function of the MC fraction. In general, the fibers with high MC fraction were flexible and exhibited remarkable toughness and high elongation at break, whereas low MC fraction lead to brittleness and overall poorer mechanical properties. In conclusion, there appears to exist an optimal composition near MC/CNC 80/20 wt-%/wt-%, where the fibers show high ductility and relatively high strength, suggested by the strain of 36.1 %, modulus of toughness value (area below the stress-strain curve) of 48.3 MJ/m$^3$ and strength 190 MPa. The modulus for the particular composition is 8.5 GPa, which suggests that the CNCs are not efficiently aligned. Notably, the plots for strain and modulus of toughness indicate that there exists a threshold value of ca. 40 wt-% for the MC fraction to initiate the plastic deformations.

To evaluate how our findings relate to the currently existing other cellulosic materials, the modulus of toughness was plotted against ultimate strain and ultimate strength together with values from other fully cellulosic and cellulose-based fibers and films (Fig. 5). In terms of toughness, MC/CNC fibers appear to be approximately on par with the top of the line cellulosic materials. This is mainly due to exceptional ductility (36.1 %), which significantly contributes to the toughness and clearly sets the MC/CNC fibers apart from the reference materials as seen in Fig. 5a, as the other reference fibers rarely exceed the ultimate strain of 15 %. On the other hand, the tensile strength of MC/CNC fibers remains in the low end of the comparison, and it appears as the weak point of the MC/CNC fibers. We still point out that the present fibers are not yet fully optimized, especially related to the aqueous concentration of MC and MC molecular weight. As a note, nanocomposite fibers consisting of, for example, silk fibroin/CNC[63] and poly(vinyl alcohol)/CNC[64] can show particularly high mechanical values, but because their main component is non-cellulosic and the fibers could be spun even without the CNC dopant, they are excluded from the comparison Fig. 5.

The effect of humidity was investigated for the MC/CNC 80.0/20.0 wt-%/wt-% fibers (Fig. S12). At low 20 % relative humidity, a reduction of mechanical properties was observed, where the ultimate strain was decreased by 28 % to 26.0 %, ultimate stress decreased by 15 % to 161.2 MPa, the modulus of toughness decreased by 38 % to 30.0 MJ/m$^3$, but Young's modulus remained nearly unchanged at 8 GPa. At high 78 % relative humidity, the ultimate strain remained roughly unchanged at 35.8 %, whereas the ultimate tensile strength severely decreased by 58 % to 79.2 MPa, modulus of toughness dropped 61 % to 18.8 MJ/m$^3$, and a drop of 52 % was detected in elastic modulus. Thus, in significantly dried conditions the ultimate strain and strength both became somewhat diminished, whereas at high humidity the ductility was well-retained but the ultimate tensile strength suffered strongly. In both cases, the modulus of toughness consequently decreased. These findings indicate the subtle role of water in plastic deformations and materials strength.

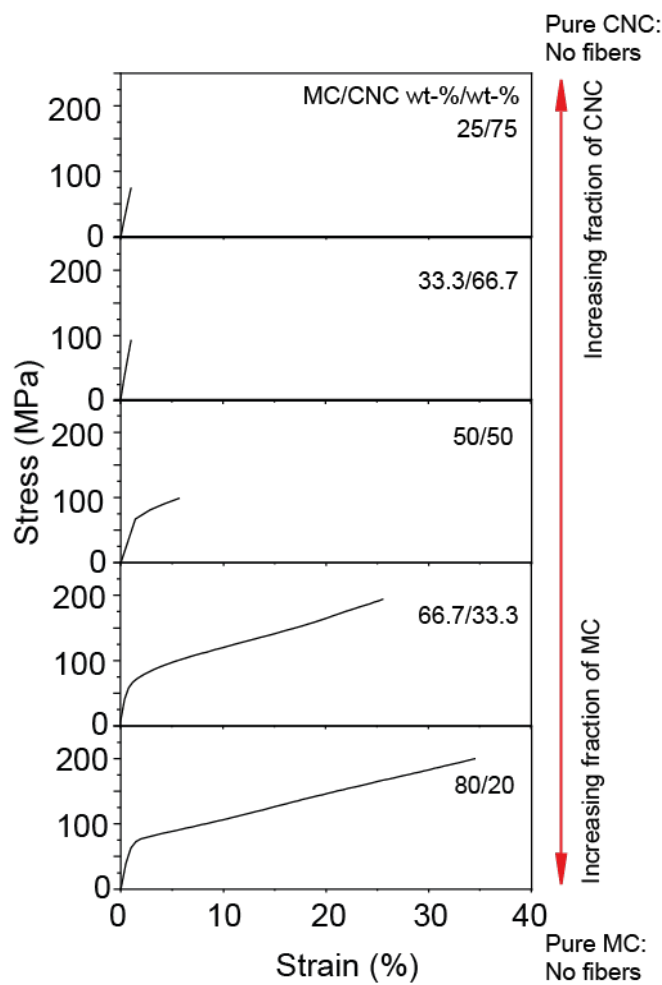

**Fig. 3.** Tensile stress-strain curves for various MC/CNC fibers spun from hydrogels, see the corresponding hydrogel compositions in Table S1. Note that neither pure CNC nor MC allowed fiber spinning.

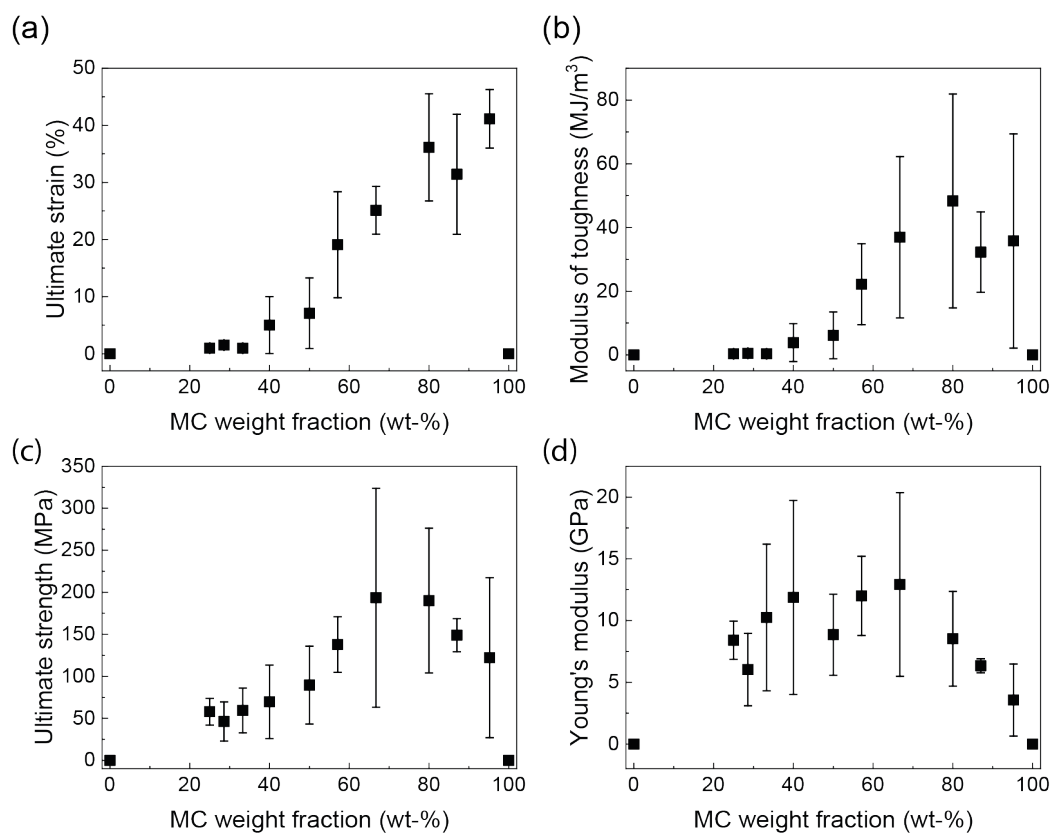

**Fig. 4.** Mechanical properties of MC/CNC fibers spun from hydrogels (see Table 1). (a) Strain; (b) modulus of toughness; (c) ultimate strength; and (d) modulus. Overall, MC/CNC 80/20 wt-%/wt-% allowed a favorable combination of ductility, strength, and toughness.

Possibilities for upscaling of the fiber production was experimented with MC/CNC 33.3/66.6 wt-%/wt-% as it allowed a robust processing in wet state in the coagulation bath (Fig. S13 and Video S2). In comparison, even though fibers with lower CNC loading produced mechanically superior fibers in the dried state, they required longer equilibration time in the coagulation bath to become solid enough to be picked up and handled and, consequently, were not suitable for large scale demonstration due to limitations of the available equipment, namely the small coagulation bath. Thus, MC/CNC 33.3/66.6 was selected due to its apparently higher wet strength. Moreover, to optimize the process, ethanol was replaced with isopropanol and the volumetric extrusion rate lowered down to 0.06 mL/min to maximize the coagulation efficiency and the time the fiber spent within the coagulation bath regardless of the constant motion. Even if post stretching has not been incorporated, the preliminary results suggest that MC/CNC fibers can be spun from aqueous gel dopes. However, to apply the method to fibers with low CNC loading, longer coagulation bath with a conveyor belt like system is recommended to allow undisturbed and long-enough coagulation time.

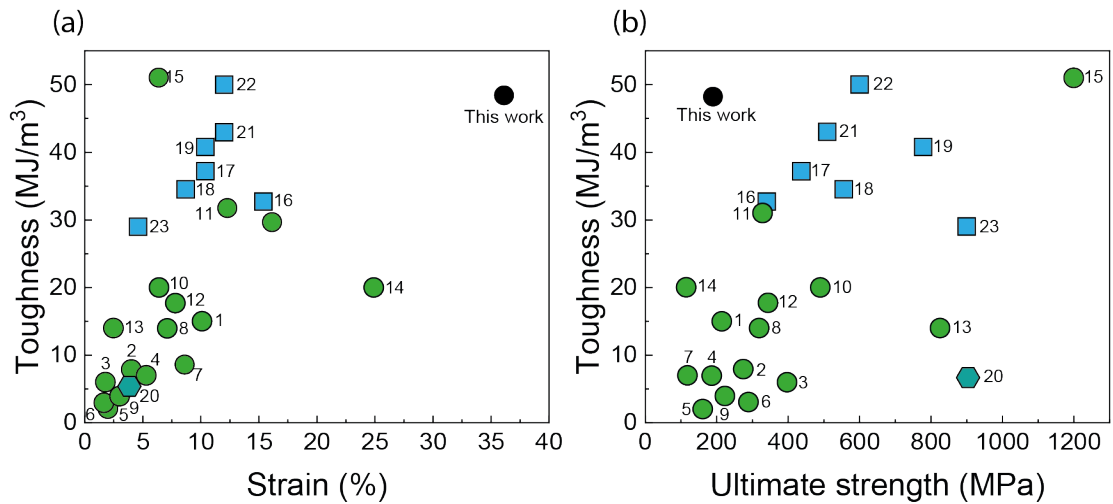

**Fig. 5.** The modulus of toughness of the MC/CNC 80/20 wt-%/wt-% fibers (black circle) plotted against (a) ultimate strain and (b) ultimate strength in comparison to values reported in the literature for all-cellulosic and cellulose-based fibers and films. Nanocellulosic reference materials are marked with green circles, commercial regenerated cellulose fibers marked with blue squares and natural flax fiber marked with turquoise hexagon. Note that in some reference materials the values for toughness were not reported and in those cases, the toughness had to be estimated based on the published stress-strain curves. The labels and reference materials are as follows: 1) Henriksson 2008[65], MFC film, 2) Walther 2011[66], CNF film, 3-4) Sehaqui 2012[67], CNF film, 5) Chen 2014[68], CA-CNC fiber, 6-7) Torres-Rendon 2014[69], CNF fiber, 8) Galland 2015[70], CNF film, 9) Hooshmand 2015[71], CNF fiber, 10) Håkansson 2014[72], CNF fiber, 11) Mohammadi 2017[54], CNF fiber, 12) Trovatti 2018[73], CNF fiber with peptide, 13-14) Wang 2017[74], Bacterial CNF fiber, 15) Mittal 2018[75], CNF fiber, 16-20) Adusumali 2006[76], Viscose, Modal, Lyocell, Rayon, and Flax fibers, respectively, and 21-23) Northolt 2001[77], Cordenka 600, Cordenka 700, and Cordenka EHM, respectively.

### 3.4 Morphology and structure of the fractured fibers

The cross-sectional morphologies of the fibers were explored by SEM after being stretched to failure. The optimal composition, i.e., MC/CNC 80/20 wt-%/wt-% showed notably clear features upon fracture, where ca. 34 nm thick rods (Fig. 6a and Fig. S14) are observed with distinct spheroidal heads, organized side-by-side in larger sheet-like structures, partially conforming to the fibers' cross-sectional shape. The length of the rods could not be reliably measured as they protruded inside the fiber interior and were closely bundled together. It is suggested that the rods consist of CNCs with MC shells, and that the spheroidal heads are formed upon restructuring of the MC shells near the fractured ends of the CNCs. The finding suggests that the CNCs are somewhat aligned along the fibrillar axis (Fig. 6d), thus contributing to the feasible mechanical properties, where the common alignment potentially allows their mutual slippage and fracture energy dissipation and, thus, ductility. Occasionally, small cracks along the fiber surface were observed by SEM in fibers after being stretched to failure indicating local structural weak points (Fig. S15). Nanocomposites with related morphologies have also been proposed for MC-CNC foams, where the bubble size, foam density, and drainage rates have been shown to be controllable by the CNC concentration.[78] At the higher CNC loadings, a different kind of packing was observed, see Fig. 6b-c for MC/CNC 40/60 wt-%/wt-% and MC/CNC 25/75 wt-%/wt-%. Especially in the latter one, the SEM micrograph hints more tightly packed and locally twisted domains, see schematically Fig. 6e. This would be expected taken the hydrogel composition MC/CNC 1.0/3.0 wt-%/wt-%, used for its preparation (see the scheme in Fig. 1f). This hydrogel composition is known to be birefringent, explained by locally twisted tactoid-like CNC aggregates, and the alignment is potentially further enhanced by the shear forces exerted during the spinning process.[43] In the intermediate composition, on the other hand, locally separated domains and areas with varying degrees of packing were observed representing the transition from smooth sheets to twisted and rigid aggregates. Accordingly, the strongly buckled and twisted macroscopic morphology of the high-CNC composite fibers with increasing CNC loadings seen in POM and SEM images (Fig. 2) is suggested to be driven by the locally twisted aggregates of the CNCs. The notion is supported by earlier observations which show that the CNC organization within CNC-alginate composite fibers is dependent on the CNC loading[79] and microphase separation of CNCs within PVA fibers[62]. Also, the poorer mechanical properties of the buckled and twisted MC/CNC fibers with less developed alignment agree with the earlier notions of the CNC-alginate and CNC-PVA composite fibers.[62,79,80] Potential concrete explanations for the weakened mechanical properties due to high CNC loading are the decreased relative amount of amorphous and flexible MC matrix and the possible nano-voids and defects within the fiber core, emerging as the side-effect of the liquid crystalline packing of the CNCs, both of which lead to fibers' inability to properly respond and restructure according to the applied external stress.

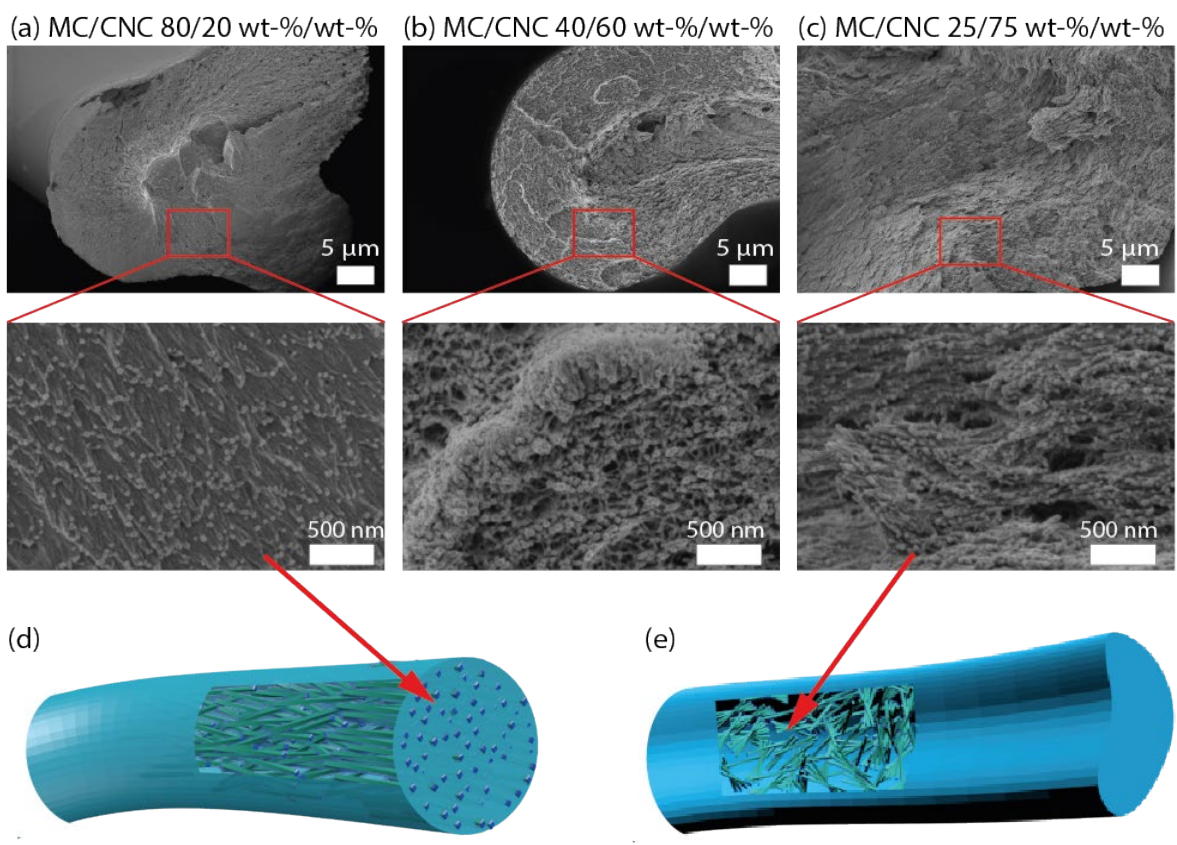

**Fig. 6.** Morphologies observed at the stretch-fractured fiber cross-sections of MC/CNC fibers with different CNC loadings. In MC/CNC 80/20 wt-%/wt-% (a), the lowest CNC concentration, rather homogenous and aligned rods were detected. In MC/CNC 40/60 wt-%/wt-% (b) and in MC/CNC 25/75 wt-%/wt-% (c) less uniformly aligned and tightly packed local structures are observed, pointing towards chiral aggregates. Schematics for fibers with low (d) and high CNC fractions (e), showing CNC alignment and twisted aggregates, correspondingly.

### 3.5 WAXS characterization

To characterize the fiber internal structural alignment, WAXS-diffractograms were measured, see Fig. 7. The studied fibers are listed in Table 2. Hermans orientation parameters were calculated from the azimuthal intensity profiles of the CNC (004) WAXS reflections (Table 2 and Fig. 7). The orientation parameters varied on an average from 0.29 for MC/CNC 80/20 to 0.42 for MC/CNC 25/75 wt-%/wt-%. Thus, the Hermans orientation parameter values remained rather low for all compositions and only a slight increase upon increasing the CNC loading was obtained. However, one explanation for the low orientation of the low-CNC loaded fibers could be their lateral spreading during the coagulation phase that could lead to the loss of spinning-derived orientation. Interestingly, the increase in orientation parameter with higher CNC loading is in contrast to the tensile tests, where the mechanical properties were found to decrease with increasing relative CNC concentration. Still, the discussion related to Figs. 3 and 6 points towards association of the CNC´s to local cholesteric tactoids with defects between them, thus reducing the overall mechanical properties. However, care must be taken in interpreting the results, since the samples with low crystalline CNC loading typically gave rather weak signals, which complicated the analysis. Recently high Hermans orientation parameters have been observed: for example, 0.7 for dry-spun CNC-CA fibers [68] and 0.96 for CNC-poly(ethylene oxide) (PEO) fibers [81] have been reported, and 0.84 for CNC/PVA-fibers.[62] These values clearly supersede the values of the MC/CNC fibers presented here. However, the ductility of MC/CNC fibers compares favorably with them.

| Compositions | | Order parameters |
|---|---|---|
| MC/CNC hydrogel solids composition (wt-%/wt-%) | MC/CNC fiber solid composition (wt-%/wt-%) | Pristine |
| 1.0/0.25 | 80.0/20.0 | $0.29 \pm 0.01$ |
| 1.0/0.5 | 66.7/33.3 | $0.32 \pm 0.02$ |
| 1.0/1.5 | 40.0/60.0 | $0.40 \pm 0.04$ |
| 1.0/3.0 | 25.0/75.0 | $0.42 \pm 0.06$ |

**Table 1.** The fiber compositions wt-%/wt-% (left column) and the observed Hermans orientation parameters of pristine un-stretched fibers.

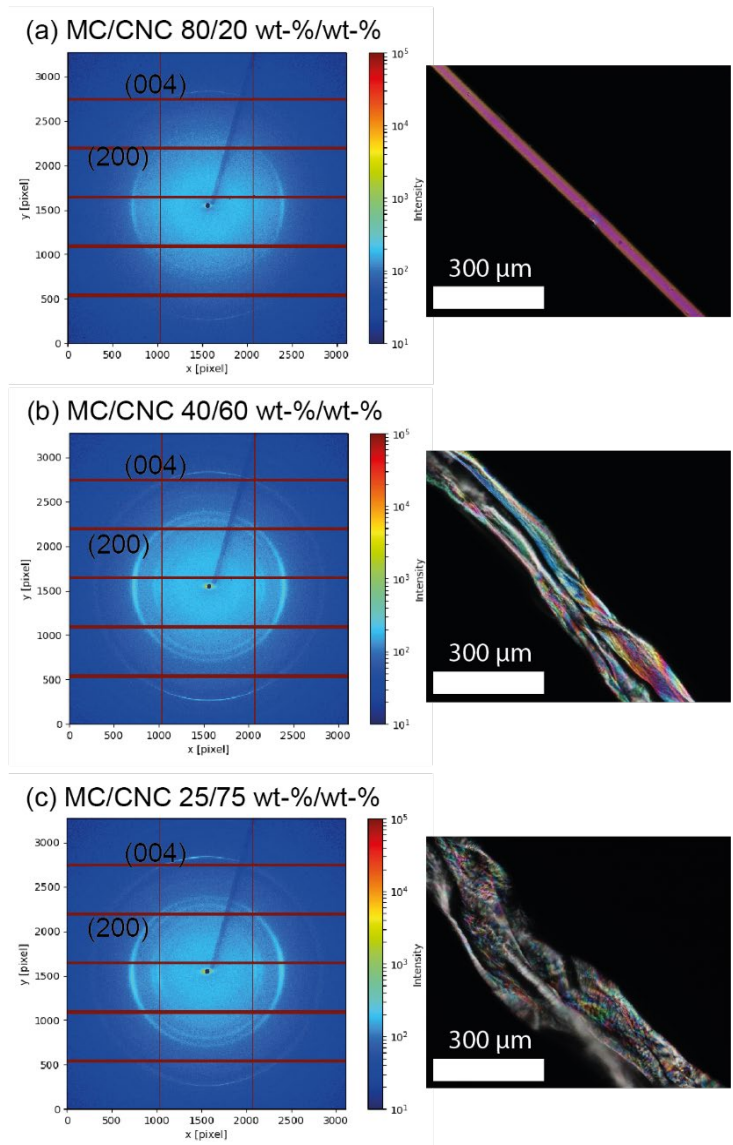

**Fig. 7.** WAXS diffractograms of MC/CNC 80/20 wt-%/wt-% (a), MC/CNC 40/60 wt-%/wt-% (b), and MC/CNC 75/25 wt-%/wt-% shown alongside respective POM images. The significant effect of CNC loading on the diffractograms and the fiber morphology is emphasized. For each fiber compositions, two fibers were measured from three different positions resulting in six data points per sample.

## 4. Conclusions

MC/CNC nanocomposite fibers possessing high ductility and modulus of toughness were prepared through a simple wet-spinning from aqueous gels into ethanol. The hydrogel MC concentration was kept fixed at 1 wt-% while the effect of CNC loading was systematically screened in the range of 0 – 3 wt-%. According to the tensile tests, mechanically the best fibers were obtained upon spinning from the hydrogel MC/CNC 1.0/0.25 wt-%/wt-%, i.e. low-end CNC fraction, leading to the fiber composition MC/CNC 80/20 wt-%/wt-%. The achieved ultimate strain and modulus of toughness values were 36.1 % and 48.3 MJ/m$^3$, respectively, while keeping a relatively high strength at 190 MPa. Higher CNC loading typically resulted in a rapid increase in brittleness. No fibers could be produced solely from MC or CNC, but their synergistic effect was required. The ductility supersedes the values up to date reported in the literature for cellulose-based nanocomposite fibers. The protocol allows to spin ductile and relatively strong methylcellulose based composite fibers from aqueous solvents.

**Acknowledgements**

This work was carried out under the Academy of Finland Centre of Excellence Programme (HYBER 2014-2019) as well as ERC for Advanced Grant (DRIVEN 2017-2022). We acknowledge the provision of facilities and technical support by Aalto University at OtaNano – Nanomicroscopy Center (Aalto-NMC). V.H. acknowledges Tekniikan Edistämissäätiö (The Finnish Foundation for Technology Promotion) for financial support.

**Appendix A. Supplementary material**

Supplementary data associated with this article can be found, in the online version, at http://dx.doi.org/xxxxxx/j.eurpolymj.